\documentclass{aa}

\usepackage{hyperref}
\usepackage[varg]{txfonts}
\usepackage{braket}
\usepackage{bm}

\renewcommand{\today}{}

\makeatletter
\renewcommand*\aa@manuscriptname{%
latex class used
\hspace{\stretch{1}}%
\copyright\ The Author \the\year
}
\makeatother

\begin{document}

\title{Upper bounds on the number of images produced by point-mass lenses
at different distances}
\titlerunning{The BKK bounds for multi-plane lensing}
\date{}
\author{J.~An\thanks{\tt jin@kasi.re.kr}}
\institute{
Korea Astronomy \& Space Science Institute,
776~Daedeok-daero, Yuseong-gu, Daejeon 34055, Korea.}
\authorrunning{An}

\abstract{If a light source is sequentially lensed by $n$ intervening point-mass lenses at distinct distances, the number of isolated images for a generic source position cannot be greater than $2\cdot3^{n-1}$. If lensing shear is present in each lens plane however, the image counts could be as large as $4^n$. For a general multi-plane point-mass lens system, we also devise the recurrence for the BKK bound on the image counts, which is strictly less than Perry's bound due to the resultant degree for $n\ge2$. None the less, this bound is still quadratic in the number of lenses on each plane, so is likely to overestimate the true sharp bound, which is expected to grow linearly with the number of lenses.}

\begingroup
\renewcommand{\today}{}
\maketitle
\endgroup
\nolinenumbers

\section{Introduction}

With the standard adoption of thin-lens approximations in gravitational lensing theory in general, multi-plane lensing still managed to attract interest among the community early on \citep[e.g.,][]{ES93}. \citet{Pe92,Pe95,Pe97}, in particular, established several foundational results regarding point-mass lenses distributed over multiple lens planes, including the minimum number of images and the difference between the numbers of even- and odd-parity images. Progress on the maximum allowed number of images, on the other hand, has been comparatively slow. For the single-plane multi-point-mass case, the problem was conclusively resolved only by \citet{KN06} although \citet{Wi90} had earlier established a weaker quadratic bound from polynomial arithmetics. As for genuine multi-plane lensing cases, \citet{Pe21} obtained an upper bound on the number of images, which essentially generalizes \citet{Wi90}. However, even for the simplest case of two point masses on two distinct planes, his bound gives 8, whereas the explicit model \citep[e.g.,][]{MWA} shows that such a system will produce either 4 or 6 images for any source not on the caustics.

This paper attempts to apply a well-established result from algebraic geometry, namely the BKK bound, to the upper bound on the number of images produced by point-mass lenses distributed over multiple distinct lens planes. Section~\ref{sec:math} outlines some mathematical preliminaries needed to formulate the problem. In Sect.~\ref{sec:olpp} and \ref{sec:mlpp}, we then explore the questions for one lens per plane and many lenses on each plane, respectively. The latter section also includes the discussion regarding the comparison of the BKK bound to that of \citet{Pe21} (Sect.~\ref{ssec:rdb}) and the effects of external shear on each lens plane (Sect.~\ref{ssec:shr}). The paper concludes with brief remarks in Sect.~\ref{sec:con}. In part due to the author's own expertise and academic background, the presentations here are based more on intuitive understanding than on full mathematical rigour, which interested readers are welcomed to be invited to provide with if so desired.

\section{Multi-plane point-mass gravitational lensing}
\label{sec:math}

Consider the propagation of light rays originating from a location on the source plane, passing through space with a given matter distribution, and arriving at an observer at the coordinate origin. Also adopted is the thin multi-plane lens approximation, in which gravitational influences on light propagation are localized to $n$ distinct planes perpendicular to the optical axis. Using the impulse approximation, the light path is modelled as $n+1$ connected straight-line segments with $n$ joints, one on each plane.

If $\bm\theta_k$ denotes the angular position vector of the impulse location on the $k$-th plane (where the observer plane is $k=0$ with $\bm\theta_0=\bm 0$, and the source plane is $k=n+1$ with source position $\bm\theta_{n+1}=\bm\theta^\mathrm S$), the multi-plane lens equations take the following form \citep{PMG}:
\begin{equation}\label{eq:leq}
\bm\theta_{k+1}=\bm\theta_1
-\sum_{j=1}^k\frac{D_{j,k+1}}{D_{k+1}}\bm\alpha_j,
\qquad(k=1,\dots,n),
\end{equation}
where $D_{j,k}$ (for $j<k$) is the angular diameter distance from the $j$-th to the $k$-th plane, $D_j=D_{0,j}$ is the angular diameter distance to the $j$-th plane from the observer, and $\bm\alpha_j$ is the angular deflection at the $j$-th plane, which is a function of the linear impulse position $D_j\bm\theta_j$ on the $j$-th plane.

In a flat background universe, the angular diameter distance is given by $D_{j,k}=(d_k-d_j)/(1+z_k)$, where $d_k$ is the comoving radial parameter to the $k$-th plane and $z_k$ is its redshift. It follows that the deflection at the $k$-th plane can be written as
\begin{equation}\label{eq:ple}
d_k\bm\alpha_k=h_{k-1,k}(\bm\theta_k-\bm\theta_{k-1})
-h_{k,k+1}(\bm\theta_{k+1}-\bm\theta_k),
\end{equation}
which is a linear function of the angular positions only on three consecutive planes including the plane of deflection and two adjacent ones \citep{Sc19,KLP}. Here $h_{j,k}=d_jd_k/(d_k-d_j)=(d_j^{-1}-d_k^{-1})^{-1}>0$ for $j<k$, and in particular, $h_{0,1}=0$. Equations (\ref{eq:leq}) and (\ref{eq:ple}) for $k=1$ are identical and no equation explicitly involves the observer position $\bm\theta_0$. In a cosmological lensing scenario with arbitrary background curvature, the three-plane lens equations can still be obtained, and in fact the derivation via the lensing time delay naturally leads to such equations \citep{Sc19}. However, the resulting expression typically relates comoving physical separations rather than observable angles \citep{PMG,Sa24}, and the coefficients scale differently. In the remainder of this paper, the assumption of a flat background is maintained, but the conclusion still stands regardless of the curvature provided that three-plane linear lens equations exist.

If the lens on the $k$-th plane can be effectively approximated as a collection of point masses, the deflection is modelled \citep{Sa24} as
\begin{equation}\label{eq:pdef}
\bm\alpha_k
=\sum_{i=1}^{m_k}\frac{4GM_{k,i}}{c^2D_k}
\frac{\bm\theta_k-\bm\lambda_{k,i}}{\|\bm\theta_k-\bm\lambda_{k,i}\|^2},
\end{equation}
where $\bm\lambda_{k,i}$ is the angular position of the lens with mass $M_{k,i}$ (on the $k$-th plane), indexed by $i\in\set{1,\dotsc,m_k}$ (assuming there are $m_k$ lenses on the $k$-th plane). If the angular unit $\vartheta=[4GM/(c^2D)]^{1/2}$ is defined by introducing a fiducial mass $M$ and a fiducial distance $D$, the lens equations with the deflections given by equation (\ref{eq:pdef}) are reducible to the complex rational equations \citep{BK75,Wi90,MWA,KLP}
\begin{gather}
\zeta_{k+1}=\zeta_1-\sum_{j=1}^k
\sum_{i=1}^{m_j}\frac{\eta_{j,k+1}^{-1}\mu_{j,i}}{\bar\zeta_j-\bar\omega_{j,i}};
\label{eq:cleq}\\\label{eq:cple}
\eta_{k-1,k}(\zeta_k-\zeta_{k-1})-\eta_{k,k+1}(\zeta_{k+1}-\zeta_k)
=\sum_{i=1}^{m_k}\frac{\mu_{k,i}}{\bar\zeta_k-\bar\omega_{k,i}}.
\end{gather}
Here $\zeta_k=(\theta_k^x+{\rm i}\theta_k^y)/\vartheta$ and $\omega_{k,j}=(\lambda_{k,j}^x+{\rm i}\lambda_{k,j}^y)/\vartheta$ are the complexified rescaled variables associated with $\bm\theta_k=(\theta_k^x,\theta_k^y)$ and $\bm\lambda_{k,j}=(\lambda_{k,j}^x,\lambda_{k,j}^y)$. The standard over-bar notation shall be used to denote the complex conjugate throughout. Also defined are $\mu_{j,i}=(1+z_j)M_{j,i}/M$ and $\eta_{j,k}=h_{j,k}/D$ to reduce clutter.

\section{BKK bounds on the number of images for lenses at distinct distances}
\label{sec:olpp}

Structurally, the lens equations define a set of mappings $\set{\bm\theta_1,\dotsc,\bm\theta_k}\mapsto\bm\theta_{k+1}$ for each $k$. In $n$-plane lensing, the lensed images for a fixed source position are represented by the vectors $\bm\theta_1$ that map to $\bm\theta^\mathrm S$ via the entire sequence of $n$ lens equations following the chain $\bm\theta_1\mapsto\bm\theta_2\mapsto\dotsc\mapsto\bm\theta_{n+1}=\bm\theta^\mathrm S$, which represents a light path terminating at the source position. If the lensing masses are all modelled as point masses, the complex lens equations (\ref{eq:cleq}) indicate that the mappings become rational functions of the complexified image positions $\zeta_j$ and their complex conjugates $\bar\zeta_j$. If $\xi_j=\bar\zeta_j$ is considered as a variable independent of $\zeta_j$ \citep{Wi90}, the set of $n$ lens equations and their complex conjugates constitutes a system of $2n$ coupled rational equations in the $2n$ variables $(\zeta_1,\xi_1,\dotsc,\zeta_n,\xi_n)$ with $\zeta_{n+1}$ and $\bar\zeta_{n+1}$ as parameters fixed by the source position. Finding lensed images for a light source deflected by point masses distributed over $n$ intervening planes at distinct distances can thus be achieved by solving for the simultaneous zeros of $2n$ polynomials (derived from the numerators of the rational functions) in $2n$ variables. Since these polynomials occur in conjugate pairs after the substitution $\bar\zeta_j\to\xi_j$, if $(\zeta_1,\xi_1,\dotsc,\zeta_n,\xi_n)$ is a solution, then $(\bar\xi_1,\bar\zeta_1,\dotsc,\bar\xi_n,\bar\zeta_n)$ is also a solution. However, only solutions for which $\bar\zeta_j=\xi_j$ for all $j=1,\dotsc,n$ actually correspond to physical image locations.

The number of isolated non-zero solutions of a system of $m$ Laurent polynomials in $m$ complex variables is known to be bounded by Bernstein's theorem \citep[also referred to as the Bernstein--Khovanskii--Kushnirenko or ``BKK'' theorem:][see also \citealt{St02}]{Be75,Ku76,Kou76,Kh78}. Since a precise statement of the theorem requires several specialized notions from algebraic and convex geometry (such as Newton polytopes, Minkowski sums, convex hulls, and mixed volumes), which are beyond the nominal scope of this paper, we shall neither reproduce the full statement nor attempt to prove the theorem here. None the less, the upper bound on the number of simultaneous zeros of a polynomial system \citep[referred to as the BKK bound, see][]{CR91} resulting from Bernstein's theorem is algorithmically computable once the set of polynomials is specified, and several public software implementations are available.

While the theorem's restriction to non-zero solutions may seem limiting at first when bounding the number of lensed images, this constraint is easily bypassed, at least for the lensing scenario of one point mass per plane ($m_k=1$ for all $k=1,\dotsc,n$). For the time being, we shall focus on this case: after defining the shifted variables $(x_k,y_k)=(\zeta_k-\omega_{k,1},\bar\zeta_k-\bar\omega_{k,1})$, the lens equations (\ref{eq:cple}) reduce to a set of Laurent polynomials (where $k=2,\dotsc,n-1$):
\begin{equation}\label{eq:lpe}\begin{split}
f_1&=x_1-x_2-q_1y_1^{-1}+c_1,\\
f_k&=a_kx_k-\eta_{k-1,k}x_{k-1}-\eta_{k,k+1}x_{k+1}-\mu_{k,1}y_k^{-1}+c_k,\\
f_n&=a_nx_n-\eta_{n-1,n}x_{n-1}-\mu_{n,1}y_n^{-1}+c_n,
\end{split}\end{equation}
in the variables $(x_1,y_1,\dotsc,x_n,y_n)$, with the shorthands
\begin{equation}\label{eq:adef}\begin{split}
q_1&=\mu_{1,1}/\eta_{1,2},\quad
c_1=\omega_{1,1}-\omega_{2,1};\quad
a_k=\eta_{k-1,k}+\eta_{k,k+1},\\
c_k&=\eta_{k-1,k}(\omega_{k,1}-\omega_{k-1,1})
+\eta_{k,k+1}(\omega_{k,1}-\omega_{k+1,1}),
\end{split}\end{equation}
while $a_n$ and $c_n$ follow the definitions of $a_k$ and $c_k$ evaluated at $k=n$ with the substitution $\omega_{n+1,1}\to\zeta_{n+1}$. An exception occurs for $n=1$, in which $f_1=x_1-q_1y_1^{-1}+c_1$ and $c_1=\omega_{1,1}-\zeta_2$. Any lensed image must correspond to a zero $(x_1,y_1,\dotsc,x_n,y_n)$ of the system of $2n$ equations $\set{f_1,\bar f_1,\dotsc,f_n,\bar f_n}$, where
\begin{equation}\label{eq:lpec}\begin{split}
\bar f_1&=y_1-y_2-q_1x_1^{-1}+\bar c_1,\\
\bar f_k&=a_ky_k-\eta_{k-1,k}y_{k-1}-\eta_{k,k+1}y_{k+1}-\mu_{k,1}x_k^{-1}+\bar c_k,\\
\bar f_n&=a_ny_n-\eta_{n-1,n}y_{n-1}-\mu_{n,1}x_n^{-1}+\bar c_n.
\end{split}\end{equation}
Here $\eta_{j,k}$ and $\mu_{k,1}$ (as well as $a_k$ and $q_1$) are all real, but the $c_k$ are complex. These equations imply that if a solution existed with $x_k=0$ or $y_k=0$ for any $k$, the source position would diverge, so such a solution could not represent a physical lensed image.

Therefore, the number of images of a source lensed by $n$ intervening point masses at distinct distances is bounded above by the BKK bound for the $2n$ Laurent polynomials $\set{f_1,\bar f_1,\dotsc,f_n,\bar f_n}$ in the $2n$ variables $(x_1,y_1,\dotsc,x_n,y_n)$ given in equations (\ref{eq:lpe}) and (\ref{eq:lpec}). In this setting, the BKK bound is the $2n$-dimensional (normalized) mixed volume of the $2n$ Newton polytopes $\set{F_1,\bar F_1,\dotsc,F_n,\bar F_n}$. Each $F_k$ is the convex hull of the origin together with four lattice points of which three have a single coordinate equal to $+1$ in the $x_{k-1}$, $x_k$, and $x_{k+1}$ directions, respectively, while the fourth has a single coordinate equal to $-1$ in the $y_k$ direction. The boundary cases $F_1$ and $F_n$ have only four vertices since the variables $x_0$ and $x_{n+1}$ are undefined (causing the corresponding vertices to collapse to the origin), whilst in the special case of $n=1$, the polytope $F_1$ is the triangle in the $(x_1,y_1)$ plane with three vertices $\set{(0,0),(1,0),(0,-1)}$. Similarly, each $\bar F_k$ is the polytope defined by the origin and four lattice points with single non-zero coordinates ($y_{k-1}=1$, $y_k=1$, $y_{k+1}=1$, or $x_k=-1$). Again, the boundary cases $\bar F_1$ and $\bar F_n$ are polytopes with four vertices, and $\bar F_n$ for $n=1$ is the triangle defined by $\set{(0,0),(0,1),(-1,0)}$.

For $n=1$, it is a simple exercise to demonstrate that the BKK bound is 2. For $n=2$ and $n=3$, we utilized SageMath's\footnote{\url{https://www.sagemath.org}} \texttt{Polyhedron} objects and \texttt{.volume()} function to compute the mixed volumes explicitly using the inclusion--exclusion formula and found the bounds to be 6 (for $n=2$; a 4-dimensional mixed volume) and 18 (for $n=3$; a 6-dimensional mixed volume). For larger values of $n$, we applied the \texttt{BKKRootBound}\footnote{\url{https://resources.wolframcloud.com/FunctionRepository/resources/BKKRootBound/}} resource function in Wolfram Mathematica. Because the version we used accepted only ordinary polynomials, we supplied $y_kf_k$ and $x_k\bar f_k$ as arguments instead. This yielded the BKK bound of $2\cdot 3^{n-1}$ for the $2n$-dimensional case up to $n=10$. These numerical values are also stricter than the bound $2(2^{2(n-1)}-1)$ derived by \citet{Pe97} for all $n>2$ since the bounds coincide for $n=2$ but the $4^n$ growth dominates over $3^n$ with increasing $n$.

Based on these findings, we propose the following:
\begin{proposition}
If a light source is sequentially lensed by $n$ intervening point-mass lenses at distinct distances, the number of isolated images for a generic source position cannot be greater than $2 \cdot 3^{n-1}$.
\end{proposition}
Here, we present a heuristic justification based on the structure of the polynomials. To this end, consider eliminating the variables $x_n$ and $y_n$ from the system $f_{n-1}=\bar f_{n-1}=f_n=\bar f_n=0$. In particular, $f_{n-1}=0$ leads to $x_n=(a_{n-1}x_{n-1}-\eta_{n-2,n-1}x_{n-2}-\mu_{n-1,1}y_{n-1}^{-1}+c_{n-1})/\eta_{n-1,n}$, and substituting this expression for $x_n$ into $f_n=0$ results in $y_n^{-1}=b_1x_{n-1}-b_2x_{n-2}-b_3y_{n-1}^{-1}+b_0$, where the $b_i$ are constants specific to the system. Then the Newton polytopes of the right-hand sides of both expressions for $x_n$ and $y_n^{-1}$ are structurally identical to $F_{n-1}$ in the $2(n-1)$-dimensional space $(x_1,y_1,\dotsc,x_{n-1},y_{n-1})$. Similarly, the equations $\bar f_{n-1}=\bar f_n=0$ provide expressions for $y_n$ and $x_n^{-1}$ as linear combinations of $y_{n-2}$, $y_{n-1}$, $x_{n-1}^{-1}$, and constants, whose Newton polytopes are structurally identical to $\bar F_{n-1}$ in $(x_1,y_1,\dotsc,x_{n-1},y_{n-1})$.

Now the variables $x_n$ and $y_n$ can be eliminated from the polynomial system by replacing $f_{n-1}=\bar f_{n-1}=f_n=\bar f_n=0$ with the equations $x_n\cdot x_n^{-1}=1$ and $y_n\cdot y_n^{-1}=1$, resulting in two polynomial equations whose Newton polytopes are both given by $F_{n-1}+\bar F_{n-1}$ in $(x_1,y_1,\dotsc,x_{n-1},y_{n-1})$. Note that the Newton polytope of a product of polynomials is the Minkowski sum of the Newton polytopes of its factors. Since each zero of the original $2n$ polynomials uniquely projects to a zero of the new, reduced system of $2n-2$ polynomials, we must have
\begin{multline}\label{eq:mvr0}
V_{2n}(F_1,\bar F_1,\dotsc,F_{n-1},\bar F_{n-1},F_n,\bar F_n)\\\le
V_{2n-2}(F_1,\bar F_1,\dotsc,F_{n-1}+\bar F_{n-1},F_{n-1}+\bar F_{n-1})
=\mathcal V_{n-1},
\end{multline}
where $V_m$ is the $m$-dimensional mixed volume of the $m$ Newton polytopes in the (unordered) argument. The inequality reflects the possibilities that variable elimination introduces spurious solutions or hidden correlations among the coefficients of the reduced system so that the particular root count may fall below the bound for the similar system with a generic set of coefficients. Multilinearity and the identity $V_{2n}(F_1,\bar F_1,\dotsc,F_n,F_n)=V_{2n}(F_1,\bar F_1,\dotsc,\bar F_n,\bar F_n)$, which follows from the symmetry under the complex conjugacy transformation $x_k\leftrightarrow y_k$, further indicate that
\begin{multline}
\mathcal V_{n-1}=2V_{2n-2}(F_1,\bar F_1,\dotsc,F_{n-1},\bar F_{n-1})
\\+2V_{2n-2}(F_1,\bar F_1,\dotsc,F_{n-1},F_{n-1}).
\end{multline}

Next, let us consider $V_{2n}(F_1,\bar F_1,\dotsc,F_n,F_n)$. The condition $f_{n-1}=f_n=0$ still yields expressions for $x_n$ and $y_n^{-1}$ as polynomials whose Newton polytope is structured like $F_{n-1}$ in $(x_1,y_1,\dotsc,x_{n-1},y_{n-1})$, while $\bar f_{n-1}=0$ results in $y_n$ being expressed as a polynomial structured like $\bar F_{n-1}$ in the same space $(x_1,y_1,\dotsc,x_{n-1},y_{n-1})$. We can then eliminate $y_n$ via $y_n\cdot y_n^{-1}=1$, whose Newton polytope is structured like $F_{n-1}+\bar F_{n-1}$ as before. However, the two copies of $F_n$ in the last layer indicate that this system actually contains two distinct polynomials of the form $b_1'x_n-b_2'x_{n-1}-b_3'y_n^{-1}+b_0'$ with generic coefficients $b_j'$. Since they are just linear equations in $x_n$ and $y_n^{-1}$, the system can be solved for $x_n$ and $y_n^{-1}$, expressed as affine functions of $x_{n-1}$ (i.e.\ linear combinations of $x_{n-1}$ and a constant). Substituting the resulting expression $x_n=\tilde b_1x_{n-1}+\tilde b_0$ into $f_{n-1}$ results in another independent polynomial, whose Newton polytope is again $F_{n-1}$ in $(x_1,y_1,\dotsc,x_{n-1},y_{n-1})$. Consequently, we can establish
\begin{multline}\label{eq:mvr1}
V_{2n}(F_1,\bar F_1,\dotsc,F_{n-1},\bar F_{n-1},F_n,F_n)
\\\phantom{V_{2n}(F_1}
\le V_{2n-2}(F_1,\bar F_1,\dotsc,F_{n-1},F_{n-1}+\bar F_{n-1})
\\\phantom{V_{2n}(F_1}
=V_{2n-2}(F_1,\bar F_1,\dotsc,F_{n-1},\bar F_{n-1})
\\+V_{2n-2}(F_1,\bar F_1,\dotsc,F_{n-1},F_{n-1})=\mathcal V_{n-1}/2,
\end{multline}
and by adding equations (\ref{eq:mvr0}) and (\ref{eq:mvr1}), we obtain
\begin{equation}
\mathcal V_n/2\le\mathcal V_{n-1}+\mathcal V_{n-1}/2\to\
\mathcal V_n\le 3\mathcal V_{n-1}\le\dotsb\le 3^{n-1}\mathcal V_1.
\end{equation}
Together with $V_2(F_1,\bar F_1)=2$ and $V_2(F_1,F_1)=1$, which are derived through explicit calculations, we have $\mathcal V_1=2[V_2(F_1,\bar F_1)+V_2(F_1,F_1)]=2(2+1)$ and $\mathcal V_n\le2\cdot 3^n$, and the conclusion follows that $V_{2n}(F_1,\bar F_1,\dotsc,F_n,\bar F_n)\le\mathcal V_{n-1}\le 2\cdot 3^{n-1}$ as conjectured.

We suspect that the mixed volume $V_{2n}(F_1,\bar F_1,\dotsc,F_n,\bar F_n)$ is actually equal to $2\cdot3^{n-1}$ (and the BKK bound is indeed $2\cdot3^{n-1}$ for all $n$). The formal proof may be constructed by establishing a recurrence for the $2n$-dimensional mixed-volume calculations via mixed-subdivision methods and applying the BKK theorem. However, we leave any attempt at a rigorous proof (or disproof if applicable) for future investigation. Nevertheless, the inequality is in fact sufficient for the proposition itself, and it explains why the computed BKK bounds follow the observed pattern.

\section{Multiple point-mass lenses per plane}
\label{sec:mlpp}

For a single plane of point-mass lenses, $\bar\zeta_1$ can be directly eliminated from the lens equation through substitution using the conjugate equation \citep{Wi90}. For a generic system of $m_1$ point masses, this results in a polynomial equation in $\zeta_1$ of degree $m_1^2+1$. For $n\ge2$, it is possible to eliminate $\set{\bar\zeta_2,\dotsc,\bar\zeta_n}$ from equation (\ref{eq:cleq}) with $k=n$ via substitutions using equations for $k=1,\dotsc,n-1$ and their conjugates \citep{MWA}. This results in a rational equation of the form $\zeta_{n+1}=\zeta_1-p(\zeta_1,\bar\zeta_1)/q(\zeta_1,\bar\zeta_1)$, where $p$ and $q$ are coprime polynomials in $(\zeta_1,\bar\zeta_1)$. The algebraic candidates for $\zeta_1$ are then among the roots of the resultant of the polynomial $p+(\zeta_{n+1}-\zeta_1)q$ and its complex conjugate with respect to $\bar\zeta_1$ with $q\ne0$. Unfortunately, this elimination rapidly becomes unwieldy, and more importantly, this procedure involving intermediate clearing of denominators can introduce many extraneous algebraic roots into the system of equations.

Instead, if each equation (\ref{eq:cple}) indexed by $k$ is multiplied by $\bar g_k=\prod_{i=1}^{m_k}(\bar\zeta_k-\bar\omega_{k,i})$, one obtains a set of $n$ polynomials
\begin{equation}\label{eq:pol}
f_k=(a_k\zeta_k-\eta_{k-1,k}\zeta_{k-1}-\eta_{k,k+1}\zeta_{k+1})\bar g_k
-\sum_{j=1}^{m_k}\mu_{k,j}\bar g_{k,j}^\mathrm s,
\end{equation}
in $2n$ variables $(\zeta_1,\bar\zeta_1,\dotsc,\zeta_n,\bar\zeta_n)$ with $\zeta_{n+1}$ fixed when $k=n$. Here $a_k=\eta_{k-1,k}+\eta_{k,k+1}$ as in equation (\ref{eq:adef}) with $a_1=\eta_{1,2}$ given $\eta_{0,1}=0$, whilst $\bar g_k$ and $\bar g^\mathrm s_{k,j}$ are univariate polynomials in $\bar\zeta_k$ defined as the complex conjugates of
\begin{equation}
g_k(\zeta_k)=\prod_{i=1}^{m_k}(\zeta_k-\omega_{k,i})
,\qquad
g^\mathrm s_{k,j}(\zeta_k)
=\prod_{\substack{i=1\\i\ne j}}^{m_k}(\zeta_k-\omega_{k,i}).
\end{equation}
Note that the degrees of $g_k$ and $g^\mathrm s_{k,j}$ are $m_k$ and $m_k-1$, respectively (if $m_k=1$, then $g^\mathrm s_{k,1}=1$, which is an empty product). The images are those common zeros of the polynomials and their conjugates ($f_k=\bar f_k=0$ for all $k=1,\dotsc,n$) that also satisfy the conjugacy condition ($\overline{\bar\zeta_k}=\zeta_k$) and avoid the lens positions ($g_k\bar g_k\ne0$).

For the single-plane ($n=1$) multi-lens case, the Newton polytope $F_1$ of $f_1$ in equation (\ref{eq:pol}) is an axis-aligned rectangle with vertices at $(0,0)$, $(0,m_1)$, $(1,m_1)$, and $(1,0)$ in the $(\zeta_1,\bar\zeta_1)$ exponent space, while the Newton polytope $\bar F_1$ of its conjugate $\bar f_1$ is the one with the $\zeta_1$ and $\bar\zeta_1$ coordinate values switched; their Minkowski sum $F_1+\bar F_1$ is a square with side length $m_1+1$. Consequently, the BKK bound on the system is $(m_1+1)^2-2m_1=m_1^2+1$, which confirms the root count obtained by direct elimination. The restriction of excluding zero solutions in the BKK bound may be formally circumvented by setting one of the lens positions to zero (e.g., $\omega_{1,m_1}=0$), which is always possible given the freedom of choice for the coordinate origin. This actually changes the structure of $f_1$ and its Newton polytope $F_1$; namely, the vertex $(1,0)$ of $F_1$ moves to $(1,1)$ (making it a right trapezium if $m_1\ge2$ or a right triangle if $m_1=1$), and the corresponding Minkowski sum results in a hexagonal shape (a square with opposite diagonal corners symmetrically truncated by right triangles) with vertices at $(0,0)$, $(0,m_1)$, $(1,m_1+1)$, $(m_1+1,m_1+1)$, $(m_1+1,1)$, and $(m_1,0)$. Nevertheless, this does not alter the BKK bound: $[(m_1+1)^2-2\cdot\frac12]-2(m_1-\frac12)=m_1^2+1$. In fact, we argue that the restriction on the non-zero solutions for the BKK bound is generally immaterial for our purposes. Since it only depends on the structures of the polynomials, the bound for a generic set of coefficients remains valid unless the system necessarily admits the zero solution. Although the system might have a solution with $\zeta_1\bar\zeta_1=0$ for a specific choice of coefficients, one can still choose the coordinate origin generically so that the solution no longer lies on the coordinate hyperplane. Given that the coordinate choice must not change the number of solutions, the BKK bound still provides an upper bound for the number of algebraic solutions to the lensing system.

However, the bound $m_1^2+1$ is sharp for the maximum allowed number of images in single-plane lensing by $m_1$ point masses only when $m_1=1$, 2, and 3 \citep{Rh01}. \citet{KN06} proved that the image count $5(m_1-1)$ attained by \citet{Rh03}'s model for $m_1\ge2$ is indeed the sharp upper bound for this case using arguments from complex dynamics \citep{KS03}. The discrepancy comes down to the fact that the BKK bound does not account for solutions failing the conjugacy constraint, which are zeros of the polynomial system but spurious solutions to the lens equation.

For $n\ge2$, the Newton polytope $F_1$ for equation (\ref{eq:pol}) is a 3-dimensional polyhedron embedded in the $2n$-dimensional exponent space with coordinates $(\zeta_1,\bar\zeta_1,\dotsc,\zeta_n,\bar\zeta_n)$, with six (or five if $m_k=1$) vertices corresponding to the monomials 1, $\zeta_1$, $\zeta_2$, $\bar\zeta_1^{m_1-1}$, $\zeta_1\bar\zeta_1^{m_1}$, $\zeta_2\bar\zeta_1^{m_1}$. The polytope $F_k$ for $f_k$ with $k\ge2$, on the other hand, is a 4-dimensional shape with eight (or seven if $m_k=1$) vertices corresponding to 1, $\zeta_{k-1}$, $\zeta_k$, $\zeta_{k+1}$, $\bar\zeta_k^{m_k-1}$, $\zeta_{k-1}\bar\zeta_k^{m_k}$, $\zeta_k\bar\zeta_k^{m_k}$, $\zeta_{k+1}\bar\zeta_k^{m_k}$ unless $k=n$, in which case $F_n$ is another 3-dimensional triangular prism with six vertices corresponding to 1, $\zeta_{n-1}$, $\zeta_n$, $\bar\zeta_n^{m_n}$, $\zeta_{n-1}\bar\zeta_n^{m_n}$, $\zeta_n\bar\zeta_n^{m_n}$. Each of these polytopes is also the Minkowski sum of the polytope with $m_k=1$ and the line segment in the $\bar\zeta_k$ direction of length $m_k-1$ (or $F_n$ is also the Minkowski sum of the triangular simplex on the $(\zeta_{n-1},\zeta_n)$-plane and the line segment in the $\bar\zeta_n$ direction of length $m_n$). In two-plane lensing ($n=2$), one can directly compute the 4-dimensional mixed volume of two 3-dimensional polytopes $F_1,F_2$ and their conjugates $\bar F_1,\bar F_2$ using these Minkowski sums (and the multilinearity of the mixed volume) and obtain the BKK bound $(m_1^2+1)(m_2^2+1)+2m_1m_2$ (Appendix~\ref{sec:app_2p}), which reduces to 6 if $m_1=m_2=1$ \citep{MWA}.

For general $n$-plane lensing, let us first consider eliminating $\zeta_n$ from the two lens equations (\ref{eq:cple}) with $k=n-1$ and $k=n$ as in the earlier case of one lens per plane. This expresses $\sum_{i=1}^{m_n}\mu_{n,i}/(\bar\zeta_n-\bar\omega_{n,i})$ as a linear combination of $\zeta_{n-2}$, $\zeta_{n-1}$, $\zeta_{n+1}$, and $\sum_{i=1}^{m_{n-1}}\mu_{n-1,i}/(\bar\zeta_{n-1}-\bar\omega_{n-1,i})$. Multiplication by $\bar g_n(\bar\zeta_n)$ first clears the denominators involving $\bar\zeta_n$, and subsequently $\bar\zeta_n$ can be eliminated by substituting the expression obtained from the conjugate equation (\ref{eq:cple}) with $k=n-1$. Once the remaining denominators involving $\zeta_{n-1}$ and $\bar\zeta_{n-1}$ are cleared via multiplication by $g_{n-1}^{m_n}\bar g_{n-1}$, the resulting polynomial is found to have a structure equivalent to $f_{n-1}\bar f_{n-1}^{m_n}$, the Newton polytope of which is reducible to the Minkowski sum $F_{n-1}+m_n\bar F_{n-1}$ (or all the monomial terms are fully contained within the corresponding convex polytope). That is to say, at the level of Newton polytopes, the last two lens planes behave as if the two pairs of polytopes $(F_{n-1},\bar F_{n-1},F_n,\bar F_n)$ were replaced by the pair of Minkowski sums $(F_{n-1}+m_n\bar F_{n-1},\bar F_{n-1}+m_nF_{n-1})$, which indicates that $V_4(F_1,\bar F_1,F_2,\bar F_2)\le V_2(F_1+m_2\bar F_1,\bar F_1+m_2F_1)=(1+m_2^2)V_2(F_1,\bar F_1)+m_2[V_2(F_1,F_1)+V_2(\bar F_1,\bar F_1)]$ for $n=2$ (where $V_{2n}$ is the $2n$-dimensional normalized mixed volume). However, $V_2(F_1,\bar F_1)=m_1^2+1$ but $V_2(F_1,F_1)=V_2(\bar F_1,\bar F_1)=2!V(F_1)$, where $V(F_1)=m_1$ is the 2-dimensional Euclidean volume (i.e.\ area) of $F_1$, so $V_4(F_1,\bar F_1,F_2,\bar F_2)\le(m_1^2+1)(m_2^2+1)+4m_1m_2$, which does not actually recover the preceding BKK bound calculation\footnote{The result does not even reduce to the earlier $m_1=m_2=1$ limit, in which one of the lens positions was set to zero. Then we have $V(F_1)=m_1-\frac12$ and recover the limit, although the result appears to be the valid BKK bound only for $m_1=1$.} $(m_1^2+1)(m_2^2+1)+2m_1m_2$, although the inequality holds.

\subsection{Recurrence for the mixed volumes and BKK bounds}

Instead, consider the decomposition of $F_n$ into the Minkowski sum of the triangle $T_n$ with the vertices corresponding to the monomials 1, $\zeta_{n-1}$, $\zeta_n$ plus the line segment $m_n\bar I_n$ with endpoints at the origin and at the exponent vector corresponding to the monomial $\bar\zeta_n^{m_n}$ (i.e.\ $\bar I_n$ is the unit line segment in the $\bar\zeta_n$-direction). From the multilinearity of the mixed volume,
\begin{multline}\label{eq:nrec}
V_{2n}(\dotsc,F_n,\bar F_n)
=V_{2n}(\dotsc,T_n+m_n\bar I_n,\bar T_n+m_nI_n)
\\=V_{2n}(\dotsc,T_n,\bar T_n)
+m_n^2V_{2n}(\dotsc,\bar I_n,I_n)
\\+m_n[V_{2n}(\dotsc,T_n,I_n)+V_{2n}(\dotsc,\bar I_n,\bar T_n)].
\end{multline}
Here, note that $T_n$ and $\bar I_n$ are the Newton polytopes of generic polynomials of the forms $\hat b_0+\hat b_1\zeta_{n-1}+\hat b_2\zeta_n$ and $\hat b_0'+\hat b_1'\bar\zeta_n$, respectively. That is to say, replacing $(F_n,\bar F_n)$ with $(T_n,\bar T_n)$ or $(\bar I_n,I_n)$ in the $2n$-dimensional polynomial system is equivalent to setting $\zeta_n$ to be either an affine function of $\zeta_{n-1}$ or a constant with the corresponding substitution for $\bar\zeta_n$. Given the structure of $F_{n-1}$ and $\bar F_{n-1}$, which are the only other polytopes involving $(\zeta_n,\bar\zeta_n)$, this procedure simply reduces the dimensionality, so
\begin{multline}\label{eq:ndef}
V_{2n}(F_1,\bar F_1,\dotsc,F_{n-1},\bar F_{n-1},T_n,\bar T_n)
\\=V_{2n}(F_1,\bar F_1,\dotsc,F_{n-1},\bar F_{n-1},\bar I_n,I_n)
\\=V_{2n-2}(F_1,\bar F_1,\dotsc,F_{n-1},\bar F_{n-1})=N_{n-1},
\end{multline}
and we find $N_n=(m_n^2+1)N_{n-1}+2m_nA_{n-1}$, where $N_n$ is the BKK bound for $n$-plane lensing and the auxiliary sequence is defined to be $A_{n-1}=V_{2n}(F_1,\bar F_1,\dotsc,T_n,I_n)=V_{2n}(F_1,\bar F_1,\dotsc,\bar I_n,\bar T_n)$.

As for the auxiliary sequence, we first observe that replacing $(\dotsc,F_n,\bar F_n)$ with $(\dotsc,T_n,I_n)$ fixes $\zeta_{n-1}$ and $\zeta_n$ to be constants but leaves $\bar\zeta_{n-1}$ and $\bar\zeta_n$ untouched. In other words, $A_n$ is the BKK bound for the system of the $n$-plane lens equations with $\zeta_n$ and $\zeta_{n+1}$ fixed but $\bar\zeta_n$ and $\bar\zeta_{n+1}$ considered as variables. For example, $A_1=V_4(F_1,\bar F_1,T_2,I_2)=V_4(F_1,\bar F_1,I_1,I_2)$ should be equal to the number of zeros of $f_1=\eta_{1,2}(\zeta_1-\zeta_2)\bar g_1(\bar\zeta_1)
-\sum_{j=1}^{m_1}\mu_{1,j}\bar g_{1,j}^\mathrm s(\bar\zeta_1)$, which is just a polynomial in $\bar\zeta_1$ of degree $m_1$ once $\zeta_1$ and $\zeta_2$ are fixed. Note that $\bar f_1$ is just a linear equation in $\bar\zeta_1$ and $\bar\zeta_2$, so $\bar\zeta_1$ uniquely specifies $\bar\zeta_2$ and $A_1=m_1$. In general, for $n\ge2$,
\begin{multline}\label{eq:arec}
A_n=V_{2n+2}(F_1,\bar F_1,\cdots,F_n,\bar F_n,T_{n+1},I_{n+1})
\\\phantom{A_n}\stackrel{(1)}{=}
V_{2n+2}(F_1,\bar F_1,\cdots,F_n,\bar F_n,I_n,I_{n+1})
\\\phantom{A_n}\stackrel{(2)}{=}
V_{2n}(F_1,\bar F_1,\cdots,T_n+m_n\bar I_n,I_n)
\\\phantom{A_n}=V_{2n}(F_1,\bar F_1,\cdots,T_n,I_n)
+m_nV_{2n}(F_1,\bar F_1,\cdots,\bar I_n,I_n)
\\\phantom{A_n}=A_{n-1}+m_nN_{n-1}.
\end{multline}
Here equality (1) expresses the fixing of $\zeta_n$ and $\zeta_{n+1}$ as constants, whereas the last two equalities follow from multilinearity and the definitions (cf.\ eq.~\ref{eq:ndef}). However, the crux is equality (2): (i) once $\zeta_{n+1}$ is considered as a constant, the polytope $F_n$ can be decomposed into the Minkowski sum in the reduced dimension; and (ii) with $\zeta_n$ fixed, $\bar f_n=0$ expresses $\bar\zeta_{n+1}$ as a linear function of $\bar\zeta_{n-1}$ and $\bar\zeta_n$, so dropping the particular polynomial does not affect the number of solutions of the system.

Based on the linear relations in equations (\ref{eq:nrec}) and (\ref{eq:arec}), we propose the following recurrence relation for the BKK bounds
\begin{equation}\label{eq:recN}\begin{split}
\begin{pmatrix}N_n\\A_n\end{pmatrix}
&=\mathsf Q_n\begin{pmatrix}N_{n-1}\\A_{n-1}\end{pmatrix}
=\mathsf Q_n\mathsf Q_{n-1}\dotsb\mathsf Q_1
\begin{pmatrix}1\\0\end{pmatrix};
\\\mathsf Q_n&=\begin{pmatrix}
m_n^2+1&2m_n\\m_n&1
\end{pmatrix},
\end{split}\end{equation}
with initial values $(N_0,A_0)=(1,0)$. Equivalently, we obtain the following second-order recurrence
\begin{equation}
m_kN_{k+1}=[(m_{k+1}^2+1)m_k+m_{k+1}]N_k+m_{k+1}(m_k^2-1)N_{k-1},
\end{equation}
with initial values $(N_0,N_1)=(1,m_1^2+1)$. Given $\det\mathsf Q_n=1-m_n^2$, if $m_n=1$, then $N_n=2A_n$, independent of $(N_{n-1},A_{n-1})$, and $N_{n+1}=(m_{n+1}^2+m_{n+1}+1)N_n$. In particular, the recurrence for $m_k=m_{k+1}=1$ reduces to the geometric progression $N_{k+1}=3N_k$, consistent with earlier findings for one point mass per plane, despite not explicitly excluding zero solutions. The recurrence formula agrees with numerical computations based on Mathematica's \texttt{BKKRootBound} in all cases that we were able to check. 

Equation (\ref{eq:recN}) also implies that $N_n$ is invariant under complete reversal of the sequence of lens multiplicities. In fact, the sequence $A_n$ may be arbitrarily rescaled while maintaining the recurrence of $N_n$. In particular,
\begin{equation}
\begin{pmatrix}N_n\\2A_n\end{pmatrix}
=\mathsf Q_n^\top\begin{pmatrix}N_{n-1}\\2A_{n-1}\end{pmatrix}
=\mathsf Q_n^\top\dotsb\mathsf Q_1^\top
\begin{pmatrix}1\\0\end{pmatrix}
=(\mathsf Q_1\dotsb\mathsf Q_n)^\top
\begin{pmatrix}1\\0\end{pmatrix},
\end{equation}
and thus the $(1,1)$-entry of the matrix $\mathsf Q_1\dotsb\mathsf Q_n$ is also $N_n$, which proves the invariance of $N_n$ under $(m_1,\dotsc,m_n)\leftrightarrow(m_n,\dotsc,m_1)$.

The general formula in equation (\ref{eq:recN}) indicates that the BKK bound grows quadratically with the number of lenses in each plane. However, the maximum number of images for the single-plane case grows linearly with the number of point masses \citep{Rh01,KN06}, and all known explicitly constructed multi-plane models have exhibited image counts whose growth is linear in each $m_k$ for fixed $n$ \citep[e.g.,][]{KLP}. That is to say, the BKK bound is expected to overcount physical images because it includes spurious solutions that fail the non-algebraic conjugacy condition.

\subsection{Bounds from resultant degrees}
\label{ssec:rdb}

For a generic $n$-plane point-mass lens system with $m_i$ lenses on each plane, \citet{Pe21} established that the number of images produced by the ensemble is bounded above by $R_n=(c_n^+)^2+(c_n^-)^2=[(\nu_n^+)^2+(\nu_n^-)^2]/2$, where $c_n^\pm=(\nu_n^+\pm\nu_n^-)/2$ are the sums of the coefficients of the even- ($+$) and odd-degree ($-$) terms in the formal polynomial $p_n=\prod_{i=1}^n(1+m_ix)$ (i.e.\ the sum of all products of an even/odd number of terms among $\set{m_1,\dotsc,m_n}$ with distinct indices) so that $\nu_n^\pm=\prod_{i=1}^n(1\pm m_i)$.\footnote{$\nu_n^+$ and $\nu_n^-$ are also the minimum number of possible images and the difference between the numbers of even- and odd-parity images, respectively \citep{Pe92} (so $c_n^\pm$ are the minimum possible numbers of even-/odd-parity images).} This bound is obtained from the degree estimate for the resultant of the two ``lensing polynomials'' obtained by eliminating the intermediate-plane variables from the lens equation (\ref{eq:cleq}) with $k=n$ and clearing the denominators.

To compare $R_n$ to our proposed BKK bound, let us observe that $p_{n+1}=(1+m_{n+1}x)p_n$ implies $c_{n+1}^\pm=c_n^\pm+m_{n+1}c_n^\mp$ and $p_0=1$ gives $(c_0^+,c_0^-)=(1,0)$. The recurrence thus follows as
\begin{equation}\label{eq:recR}
\begin{pmatrix}
R_{n+1}\\c_{n+1}^+c_{n+1}^-
\end{pmatrix}=\begin{pmatrix}
1+m_{n+1}^2&4m_{n+1}\\m_{n+1}&1+m_{n+1}^2
\end{pmatrix}\begin{pmatrix}
R_n\\c_n^+c_n^-
\end{pmatrix},
\end{equation}
with initial values $R_0=(c_0^+)^2+(c_0^-)^2=1$ and $c_0^+c_0^-=0$.
Since $m_{n+1}\ge1$, all the matrix elements here are greater than or equal to the corresponding entries in $\mathsf Q_{n+1}$ in equation (\ref{eq:recN}), so $N_n\le R_n$ for $n\ge0$ and $N_n<R_n$ for $n\ge2$.

The improvement of the BKK bound for the unreduced $2n$-dimensional system over the resultant-degree bound may be understood from the sparsity of the lensing polynomials (i.e.\ the loss of support information in the degree-based bound) and their non-genericity arising from eliminating intermediate-layer variables. Note that the resultant theorem used by \citet[theorem~2]{Pe21} also follows from the BKK theorem and monotonicity of mixed volumes. In particular, if $p(x,y)$ and $q(x,y)$ are coprime bivariate polynomials, their Newton polygons $P$ and $Q$ are contained in the axis-aligned rectangles $\mathcal R_p=[0,p_x]\times[0,p_y]$ and $\mathcal R_q=[0,q_x]\times[0,q_y]$ with $p_x=\deg_xp$ etc.\ (i.e.\ $P\subseteq\mathcal R_p$ and $Q\subseteq\mathcal R_q$). Monotonicity of mixed volumes then implies $V_2(P,Q)\le V_2(\mathcal R_p,\mathcal R_q)=(p_x+q_x)(p_y+q_y)-p_xp_y-q_xq_y=p_xq_y+q_xp_y$. This resultant bound concerns only the degrees in each variable and discards information about the Newton polygons, whose mixed area may be smaller. However, this limitation applies to the degree estimate, rather than resultant methods in general. In principle, sparse resultants can retain the support information \citep{DS15}.

Nevertheless, $R_2=(1+m_1m_2)^2+(m_1+m_2)^2=(m_1^2+1)(m_2^2+1)+4m_1m_2$ is consistent with the earlier computation of $V_2(F_1+m_2\bar F_1,\bar F_1+m_2F_1)$ based on eliminating variables. In fact, the Newton polytope of the two-plane lensing polynomial is the axis-aligned rectangle $[0,m_1m_2+1]\times[0,m_1+m_2]$ assuming generic locations\footnote{If one of the lens locations on the first plane is set to the coordinate origin (e.g., $\omega_{1,m_1}=0$, which is always possible without loss of generality among represented models), the coefficients of $\zeta_1^{m_1m_2+1}$ and $\bar\zeta_1^{m_1+m_2}$ in the lensing polynomial vanish and the corresponding Newton polytope becomes a hexagon resulting from the corner-cut rectangle while its mixed volume with the conjugate is $(1+m_1m_2)^2+(m_1+m_2)^2-2m_2$.} for the lenses on both planes and the source. However, the direct four-dimensional computation in Appendix~\ref{sec:app_2p} results in the strictly smaller bound $V_4(F_1,\bar F_1,F_2,\bar F_2)=(m_1^2+1)(m_2^2+1)+2m_1m_2$, which cannot be explained by missing support information since those rectangles are already the Newton polytopes.
Moreover, the outlined elimination procedure also indicates that the two-dimensional Newton polygon $\mathcal F_n$ ($n\ge2$) of the reduced lensing polynomial for the $n$-plane system is given by the Minkowski sum $\mathcal F_n=\mathcal F_{n-1}+m_n\overline{\mathcal F}_{\!n-1}$, so $V_2(\mathcal F_n,\overline{\mathcal F}_{\!n})=(1+m_n^2)V_2(\mathcal F_{n-1},\overline{\mathcal F}_{\!n-1})+2m_nV_2(\mathcal F_{n-1},\mathcal F_{n-1})$ and $V_2(\mathcal F_n,\mathcal F_n)=(1+m_n^2)V_2(\mathcal F_{n-1},\mathcal F_{n-1})+2m_nV_2(\mathcal F_{n-1},\overline{\mathcal F}_{\!n-1})$. Consequently, if $V_2(\mathcal F_n,\mathcal F_n)=2!V(\mathcal F_n)=2c_n^+c_n^-$ (which is consistent with $\mathcal F_n$ being the axis-aligned rectangle $[0,c_n^+]\times[0,c_n^-]$), then the mixed volume $V_2(\mathcal F_n,\overline{\mathcal F}_{\!n})$ follows the recurrence for $R_n$ in equation (\ref{eq:recR}).

We observe not only that the classical resultants may introduce spurious solutions (intersections at infinity separate from those associated with clearing denominators) into the system but also that elimination imposes algebraic relations among the coefficients of the reduced polynomials. For instance, in the simplest multi-plane case of $n=2$ and $m_1=m_2=1$, the resultant of the reduced lensing polynomial and its complex conjugate with respect to $\bar\zeta$ is a degree-seven polynomial\footnote{The resultant sextic in \citet{MWA} is actually a further reduced form with $m_1z$ (in their notation) factored out.} in $\zeta$ whereas the resultant of generic polynomials on the same support is in fact a polynomial of degree eight. That is to say, the coefficient of the highest-degree term in a generic case is such that it is identically zero with the particular choices of coefficients for the lensing model. This particular reduction of the degree appears to be related to the freedom of choice for the coordinate origin -- if one sets $\omega_{1,1}=0$ (which is physically immaterial), the coefficients of $\zeta^2$ and $\bar\zeta^2$ in the reduced lensing polynomial vanish and the generic counterpart of the resulting resultant is a polynomial of degree seven. In addition, the degree-seven resultant of the lensing polynomials actually has a factor $(\zeta_1-\omega_{1,1})$ which is due to clearing the denominators. That is to say, the resultant may contain a factor of the denominators in the original rational equations as one of its own factors. If these hidden restrictions are ignored, the root bound associated with the structures of the reduced polynomials for generic coefficients need not be attained by the particular reduced system. A more precise analysis would compare a resultant obtained after the intermediate substitutions with a polynomial obtained by eliminating all but the same variable from the full $2n$-equation system (also excluding those positions ending up on the lenses on the intermediate layers). The remaining question is how the specialized lensing coefficients lead to the correlated coefficients in the resultants and reduced polynomials and affect the sharpness of the resulting bounds on the root counts, but we leave this analysis for the future.

\subsection{External shear}
\label{ssec:shr}

If significant tidal fields from external mass distributions are present on each lens plane, their leading-order effects can be approximated by adding a quadratic term in position to the lensing potential. If the local mass column densities vanish except at the point masses, the tidal contribution to the lensing potential is represented by a harmonic function, and in terms of the complex lens equation, this introduces (complex) lensing shear to the deflections \citep{AE06}. In particular, for multiple point masses on each lens plane, equations (\ref{eq:cple}) become
\begin{equation}\label{eq:shr}
\sum_{i=1}^{m_k}\frac{\mu_{k,i}}{\bar\zeta_k-\bar\omega_{k,i}}
+\gamma_k\bar\zeta_k
=\eta_{k,k+1}(\zeta_k-\zeta_{k+1})-\eta_{k-1,k}(\zeta_{k-1}-\zeta_k),
\end{equation}
where $\gamma_k$ is the complex representation of the symmetric, traceless rank-two tensor describing the tidal shear (which relates to the second-order Wirtinger derivatives of the harmonic lensing potential due to the tidal effects). The scales are chosen so that the distance factors in each plane match those of the point-mass deflection terms. There is also a sign ambiguity relating to the definition of the ``direction'' of the shear, but the details are immaterial for our purposes here.

For algebraic root counting, multiplying each equation (\ref{eq:shr}) with index $k$ by the same polynomial $\bar g_k(\bar\zeta_k)$ as in the null-shear case gives a set of $n$ polynomial equations in the same $2n$ variables. For $n=1$, the Newton polytope of the lens equation is the quadrilateral with vertices at $(0,0)$, $(1,0)$, $(1,m_1)$, and $(0,m_1+1)$ (again a trapezium), whilst the Minkowski sum of this polytope with that of the conjugate equation is the pentagon with vertices $(0,0)$, $(0,m_1+2)$, $(m_1,m_1+2)$, $(m_1+2,m_1)$, $(m_1+2,0)$ (i.e.\ a square of side length $m_1+2$ with its top-right corner truncated by a right isosceles triangle of base length 2). Consequently, the BKK bound is $[(m_1+2)^2-2]-2(m_1+1/2)=(m_1+1)^2$, which again agrees with the root count obtained by direct elimination. More interestingly, the degree of the polynomial equation obtained after multiplying the lens equation by $\bar g_1$ is $m_1+1$, so the BKK bound coincides with the B\'ezout bound. In fact, for general $n$-plane lensing with external shear on each plane, we observe that the BKK bound on the algebraic roots of the set of $n$ lens equations (\ref{eq:shr}) and their conjugates reproduces the B\'ezout bound $\prod_{j=1}^n(m_j+1)^2$ (Appendix~\ref{sec:app_sh}). If each lens plane contains a single point mass ($m_j=1$), this bound reduces to $4^n$.

As in the null-shear case, however, this algebraic bound is expected to overcount physical images once the conjugacy constraints are imposed. For example, in the single-plane case, complex-dynamical arguments give the stronger physical upper bound $5m_1-1$ for the number of images in the presence of external shear \citep{AE06}. The algebraic bound $(m_1+1)^2$ is sharp for the physical image count only for $m_1=1$ ($N_1\le4$) and $m_1=2$ ($N_1\le9$). Although we expect that the bound $4^n$ may conceivably be sharp when there is one point mass on each lens plane and external shear is present, it is quite likely that the conjugacy constraints again reduce the sharp physical bound on the image count below the algebraic (BKK or B\'ezout) bounds. Establishing the corresponding sharp physical bound, however, requires further arguments beyond the algebraic root counting.

\section{Concluding remarks}
\label{sec:con}

The main conclusions of this paper are that (i) the number of gravitationally lensed images of a source produced by point masses distributed over $n$ separate planes is bounded above by the BKK bound of the $2n$-variable polynomial system obtained by clearing the denominators of the lens equations; (ii) this bound $N_n$ satisfies the recurrence in equation~(\ref{eq:recN}), where $m_k$ is the number of point masses on the $k$-th plane; and (iii) it reduces to $2\cdot3^{n-1}$ when all $m_k=1$. With external shear on each plane, the bound becomes $\prod_{k=1}^n(m_k+1)^2$, or $4^n$ when all $m_k=1$. We suspect that the bound may plausibly be sharp for any $n$ when there is only one point mass per plane. For comparison, the sharp lower bound in the shear-free case is $2^n$, attained by collinearly aligned $n$ point masses for an off-axis source. However, it is doubtful that the bound for many lenses on each plane is close to the actual sharp physical upper bound once the conjugacy condition is introduced into the algebraic solution sets. This is already evident in the single-plane multi-lens example, in which the sharp upper bound grows linearly with the number of lenses, whereas the algebraic bound is quadratic.

It is then instructive to examine the known basis of this linear bound. Let us observe that the multi-plane point-mass lens equation can also be written as
\begin{equation}
\zeta_k=\zeta_{n+1}
+\sum_{j=1}^n\frac{\overline{\delta_j(\zeta_j)}}{\eta_{p,n+1}}
;\quad
\delta_j(x)=\sum_{i=1}^{m_j}\frac{\mu_{j,i}}{x-\omega_{j,i}}+\bar\gamma_jx,
\end{equation}
where $p=\max(j,k)$ and $k=1,\dotsc,n$. We define mappings $\mathsf h,\bar{\mathsf h}:\widehat{\mathbb C}^n\to\widehat{\mathbb C}^n$ where $\widehat{\mathbb C}^n$ is the direct product of $n$ Riemann spheres $\widehat{\mathbb C}=\mathbb C\cup\set{\infty}$, wherever the following expressions are defined:
\begin{equation}
\mathsf h(x_1,\dotsc,x_n)=(h_1,\dotsc,h_n);\quad
\bar{\mathsf h}(x_1,\dotsc,x_n)
=(\bar h_1,\dotsc,\bar h_n),
\end{equation}
with
\begin{equation}\begin{split}
h_k(x_1,\dotsc,x_n)&=\bar\zeta_{n+1}
+\sum_{j=1}^n\frac{\delta_j(x_j)}{\eta_{p,n+1}},
\\\bar h_k(x_1,\dotsc,x_n)&=\zeta_{n+1}
+\sum_{j=1}^n\frac{\bar\delta_j(x_j)}{\eta_{p,n+1}};
\\\bar\delta_j(x)&=\overline{\delta_j(\bar x)}=\sum_{i=1}^{m_j}
\frac{\mu_{j,i}}{x-\bar\omega_{j,i}}+\gamma_jx.
\end{split}\end{equation}
At finite positions away from the lenses, the lens equation and its complex conjugate are equivalent to $\mathsf h(\zeta_1,\dotsc,\zeta_n)=(\bar\zeta_1,\dotsc,\bar\zeta_n)$ and $\bar{\mathsf h}(\bar\zeta_1,\dotsc,\bar\zeta_n)=(\zeta_1,\dotsc,\zeta_n)$. In other words, physical image positions $(\zeta_1,\dotsc,\zeta_n)$ for a fixed source position $\zeta_{n+1}$ are also among the fixed points of the rational map $\bar{\mathsf h}\circ\mathsf h$. Technically, if $\bm\zeta\in\widehat{\mathbb C}^n$ is a fixed point of $\bar{\mathsf h}\circ\mathsf h$, then $\mathsf h(\bm\zeta)=\bm\xi$ is a fixed point of $\mathsf h\circ\bar{\mathsf h}$ and $\bar{\mathsf h}(\bm\xi)=\bm\zeta$. However, the fixed point $\bm\zeta$ corresponds to a physical lensed image only if $\bm\xi=\bar{\bm\zeta}$.

This approach is useful for $n=1$ because attracting fixed points of a rational function of one complex variable are linked to its critical points, the fact upon which \citet{KN06} built their proof of the upper bound of $5(m_1-1)$ on the number of images for single-plane multi-point-mass lensing without shear when $m_1\ge2$ \citep[see also][]{KS03,AE06}. An important nuance here is that the number of even-parity images is bounded not merely by the critical-point count of the composite map $\bar{\mathsf h}\circ\mathsf h$, but by that of the constituent map $\mathsf h$, which grows linearly with its degree and hence with the number of deflectors.

For $n\ge2$, however, this one-variable argument does not extend directly, as there are no immediate analogues to the connection between attracting fixed points and critical points nor the critical-point counting formula (the Riemann–Hurwitz formula) known in the complex dynamics of several variables. Instead, a genuinely several-variable complex-dynamical approach is likely to be necessary to turn this fixed-point formulation into a generic bound on the image count. Nevertheless, a subset of images (at least all those corresponding to attracting fixed points whose basins of attraction contain the chosen initial conditions---can be located by alternating the two rational maps, with conjugacy checked at convergence, rather than solving for the simultaneous roots of multiple coupled rational functions and individually checking them for conjugacy.

\begin{acknowledgements}
This research has been supported financially by the internal KASI research grant (project no.\ 2026183200).
\end{acknowledgements}

\begin{appendix}
\nolinenumbers

\section{BKK bounds for two-plane point-mass lensing}
\label{sec:app_2p}

In this appendix, we compute $N_2=V_4(F_1,\bar F_1,F_2,\bar F_2)$ directly. 

First, let us observe that
\begin{equation}
f_1=\eta_{1,2}(\zeta_1-\zeta_2)\bar g_1-\sum_{j=1}^{m_1}\mu_{1,j}\bar g^\mathrm s_{1,j}
\end{equation}
given $\eta_{0,1}=0$ and $a_1=\eta_{0,1}+\eta_{1,2}=\eta_{1,2}$. Here
\begin{equation}
\bar g_1=\prod_{i=1}^{m_1}(\bar\zeta_1-\bar\omega_{1,i});\qquad
\bar g^\mathrm s_{1,j}=\prod_{\substack{i=1\\i\ne j}}^{m_1}(\bar\zeta_1-\bar\omega_{1,i}),
\end{equation}
which are polynomials in $\bar\zeta_1$ of degrees $m_1$ and $m_1-1$, respectively. Then $f_1$ contains the monomials $\zeta_2\bar\zeta_1^a$, $\zeta_1\bar\zeta_1^a$, and $\bar\zeta_1^b$ with $a=0,\dotsc,m_1$ and $b=0,\dotsc,m_1-1$ for generic lens positions $\omega_{1,i}$. Since the Newton polytope of a polynomial is the convex hull of the exponent vectors of the monomials with non-zero coefficients, the Newton polytope $F_1$ of $f_1$ in the four-dimensional exponent space associated with $(\zeta_1,\bar\zeta_1,\zeta_2,\bar\zeta_2)$ is a three-dimensional polyhedron (with zero extent in the $\bar\zeta_2$-direction) with vertices at $(0,0,0,0)$, $(0,m_1-1,0,0)$, $(1,0,0,0)$, $(1,m_1,0,0)$, $(0,0,1,0)$, and $(0,m_1,1,0)$. If $F^0$ is the Newton polytope $F_1$ for $m_1=1$, i.e.\ the convex polyhedron with vertices at $(0,0,0,0)$, $(1,0,0,0)$, $(0,0,1,0)$, $(1,1,0,0)$, and $(0,1,1,0)$, then $F_1$ for any $m_1\ge1$ corresponds to the region swept out by translating $F^0$ through distances from 0 to $m_1-1$ along the $\bar\zeta_1$-coordinate direction; that is, $F_1$ is the Minkowski sum $F_1=F^0+(m_1-1)\bar I_1$ where $\bar I_1$ is the line segment of unit length in the $\bar\zeta_1$-direction.
On the other hand, the polynomial $f_2$ consists of the monomials $\zeta_1\bar\zeta_2^a$, $\zeta_2\bar\zeta_2^a$, $\zeta_3\bar\zeta_2^a$, and $\bar\zeta_2^b$ with $a=0,\dotsc,m_2$ and $b=0,\dotsc,m_2-1$. However, since $\zeta_3$ is considered as a constant parameter in the same four-dimensional exponent space, the Newton polytope $F_2$ is a prism with the triangular base $T_2$ whose vertices are at $(0,0,0,0)$, $(1,0,0,0)$, and $(0,0,1,0)$, swept in the $\bar\zeta_2$-direction from 0 to $m_2$; that is, $F_2=T_2+m_2\bar I_2$.
By multilinearity of the mixed volumes, $N_2$ can be expressed as a linear combination of 16 four-dimensional mixed volumes of the polytopes with fixed vertices independent of $m_1$ and $m_2$, all of which can be computed using software for mixed-volume computation. 

Alternatively,
\begin{multline}
N_2=V_4(F_1,\bar F_1,T_2+m_2\bar I_2,\bar T_2+m_2I_2)
\\\phantom{N_2}
=V_4(F_1,\bar F_1,T_2,\bar T_2)+m_2^2V_4(F_1,\bar F_1,I_2,\bar I_2)
\\+m_2[V_4(F_1,\bar F_1,T_2,I_2)+V_4(F_1,\bar F_1,\bar I_2,\bar T_2)].
\end{multline}
Here we introduce a projection identity for mixed volumes with a unit coordinate segment as one argument; namely,
\begin{multline}
V_{m+1}(P_1,\dotsc,P_m,E_i)
=V_{m+1}(\pi_iP_1,\dotsc,\pi_iP_m,E_i)
\\=V_m(\pi_iP_1,\dotsc,\pi_iP_m),
\end{multline}
where $E_i$ is the unit line segment in the $i$-th coordinate direction, and $\pi_iP$ is the projection of the polytope $P$ onto the coordinate hyperplane perpendicular to the $i$-th axis (i.e., the polytope resulting from collapsing all its $i$-th coordinate values to zero). This follows as a corollary to Lemma~4 of \citet{ST10}. Note then that $V_4(F_1,\bar F_1,I_2,\bar I_2)=V_2(\pi F_1,\pi\bar F_1)=N_1$, where $\pi F_1=[0,1]\times[0,m_1]=I_1+m_1\bar I_1$ is the projection of $F_1$ to the two-dimensional $(\zeta_1,\bar\zeta_1)$-space and $\pi\bar F_1=\bar I_1+m_1I_1$. Thus $N_1=m_1^2+1$ given $V_2(I_1,\bar I_1)=1$ and $V_2(I_1,I_1)=V_2(\bar I_1,\bar I_1)=0$.

Next, the projection of $F^0$ onto the $(\bar\zeta_1,\bar\zeta_2)$-plane is the unit line segment $\bar I_1$ in the $\bar\zeta_1$ direction, whereas that of $\bar F^0$ is the triangle $\bar T_2$. Since $F_1=F^0+(m_1-1)\bar I_1$, we then find
\begin{multline}
A_1=V_4(F_1,\bar F_1,T_2,I_2)=V_4(F_1,\bar F_1,I_1,I_2)
\\=V_4(m_1\bar I_1,\bar T_2,I_1,I_2)
=m_1V_4(\bar I_1,\bar I_2,I_1,I_2)=m_1.
\end{multline}
In addition, the symmetry of the system under interchanging the barred and unbarred exponent coordinates implies $V_4(F_1,\bar F_1,\bar I_2,\bar T_2)=V_4(F_1,\bar F_1,T_2,I_2)=A_1=m_1$.

Lastly,
\begin{multline}
V_4(F_1,\bar F_1,T_2,\bar T_2)
\\=V_4(F^0+(m_1-1)\bar I_1,\bar F^0+(m_1-1)I_1,T_2,\bar T_2)
\\=V_4(F^0,\bar F^0,T_2,\bar T_2)
+(m_1-1)^2V_4(\bar I_1,I_1,T_2,\bar T_2)
\\+(m_1-1)[V_4(F^0,I_1,T_2,\bar T_2)
+V_4(\bar I_1,\bar F^0,T_2,\bar T_2)],
\end{multline}
but successive projections simplify the terms to
$V_4(\bar I_1,I_1,T_2,\bar T_2)=V_4(\bar I_1,I_1,I_2,\bar I_2)=1$ and
$V_4(F^0,I_1,T_2,\bar T_2)=V_4(\bar I_1,I_1,I_2,\bar T_2)=V_4(\bar I_1,I_1,I_2,\bar I_2)=1$.

As for the last remaining term, we first observe that the monotonicity of mixed volumes indicates, given $F^0\subset T_2+\bar I_1$ and $I_2\subset T_2$,
\begin{multline}
V_4(F^0,\bar F^0,I_2,\bar I_2)\le
V_4(F^0,\bar F^0,T_2,\bar T_2)\\\le
V_4(T_2+\bar I_1,\bar T_2+I_1,T_2,\bar T_2)
\\=V_4(T_2,\bar T_2,T_2,\bar T_2)+V_4(\bar I_1,I_1,T_2,\bar T_2)
\\+V_4(T_2,I_1,T_2,\bar T_2)
+V_4(\bar I_1,\bar T_2,T_2,\bar T_2).
\end{multline}
Now $V_4(F^0,\bar F^0,I_2,\bar I_2)=2$ since this is the same as $N_1$ for $m_1=1$. Next $V_4(T_2,I_1,T_2,\bar T_2)=V_4(I_2,I_1,I_2,\bar T_2)=0$ and so is $V_4(\bar I_1,\bar T_2,T_2,\bar T_2)=0$, whilst $V_4(\bar I_1,I_1,T_2,\bar T_2)=V_4(\bar I_1,I_1,I_2,\bar I_2)=1$. Finally, \citet[Lemma~4]{ST10} indicate
\begin{equation}
V_4(T_2,\bar T_2,T_2,\bar T_2)=V_2(T_2,T_2)V_2(\bar T_2,\bar T_2)
\end{equation}
where $V_2(T_2,T_2)$ is the 2-dimensional mixed volume in the $(\zeta_1,\zeta_2)$-space and $V_2(\bar T_2,\bar T_2)$ is the same in the $(\bar\zeta_1,\bar\zeta_2)$-space. The triangle $T_2$ has Euclidean area $V(T_2)=1/2$, so its BKK-normalized mixed area with itself is $V_2(T_2,T_2)=2!V(T_2)=1$ and $V_2(\bar T_2,\bar T_2)=1$ by symmetry. Therefore, we find that $V_4(T_2,\bar T_2,T_2,\bar T_2)=1$ and $V_4(F^0,\bar F^0,T_2,\bar T_2)=2$. Combining all these results, we obtain $V_4(F_1,\bar F_1,T_2,\bar T_2)=2+(m_1-1)^2+2(m_1-1)=m_1^2+1$ and $N_2=(1+m_2^2)(1+m_1^2)+2m_2m_1$.

\section{Recurrence for BKK bounds with shear}
\label{sec:app_sh}

The polynomial obtained by multiplying equation (\ref{eq:shr}) by $\bar g_k(\bar\zeta_k)$ gains additional monomials $\bar\zeta_k^{1+a}$ with $a=0,\dotsc,m_k$ compared to $f_k$ in equation (\ref{eq:pol}), so the corresponding polytope $F_k$ has the vertex corresponding to the monomial $\bar\zeta_k^{m_k+1}$ instead of $\bar\zeta_k^{m_k-1}$ (or $\bar\zeta_n^{m_n}$ if $k=n$). Then the polytope $F_n$ in the $2n$-dimensional space $(\dotsc,\zeta_n,\bar\zeta_n)$ is the Minkowski sum $\Delta_n+m_n\bar I_n$, where $\Delta_n$ is a tetrahedron with vertices corresponding to the monomials 1, $\zeta_{n-1}$, $\zeta_n$, $\bar\zeta_n$, so
\begin{multline}
V_{2n}(\dotsc,F_n,\bar F_n)
=V_{2n}(\dotsc,\Delta_n+m_n\bar I_n,\bar\Delta_n+m_nI_n)
\\=V_{2n}(\dotsc,\Delta_n,\bar\Delta_n)
+m_n^2V_{2n}(\dotsc,I_n,\bar I_n)
\\+m_n[V_{2n}(\dotsc,\Delta_n,I_n)+V_{2n}(\dotsc,\bar\Delta_n,\bar I_n)].
\end{multline}
Here $V_{2n}(\dotsc,I_n,\bar I_n)$ is again the generic root count for the system while fixing both $\zeta_n$ and $\bar\zeta_n$, i.e.\ the BKK bound $N_{n-1}$ corresponding to the reduced dimension. On the other hand, the mixed volume $V_{2n}(\dotsc,\Delta_n,\bar\Delta_n)$ corresponds to the root count of the system with the last two equations replaced by two generic linear equations of the form $b_0+b_1\zeta_{n-1}+b_2\zeta_n+b_3\bar\zeta_n=0$ and its complex conjugate. These two equations can generically be solved for $(\zeta_n,\bar\zeta_n)$ given as affine functions of $\zeta_{n-1}$ and $\bar\zeta_{n-1}$, so this reduces to the root count of the $2(n-1)$ dimensional system with $(\zeta_n,\bar\zeta_n)$ in $(f_{n-1},\bar f_{n-1})$ replaced by a linear combination $\zeta_n=\hat b_0+\hat b_1\zeta_{n-1}+\tilde b_1\bar\zeta_{n-1}$ and its conjugate. The monomials $\zeta_n\bar\zeta_{n-1}^a$ in $f_{n-1}$ then result in $\bar\zeta_{n-1}^a$, $\zeta_{n-1}\bar\zeta_{n-1}^a$ and $\bar\zeta_{n-1}^{a+1}$, so the support of the reduced polynomial in $2(n-1)$ dimensions is actually identical to the projection of the $2n$-dimensional support of $f_{n-1}$ collapsed onto the space of constant $\zeta_n$ (the vertex term $\bar\zeta_{n-1}^{1+m_{n-1}}$ was already present in the support of $f_{n-1}$ with shear). Together with a symmetric argument for $\bar f_{n-1}$, we can conclude $V_{2n}(\dotsc,\Delta_n,\bar\Delta_n)=V_{2n}(\dotsc,I_n,\bar I_n)$.

As for the mixed volume $V_{2n}(\dotsc,\Delta_n,I_n)$, this corresponds to the generic root count of the system with the last two equations, $b_0+b_1\zeta_{n-1}+b_2\zeta_n+b_3\bar\zeta_n=0$ and $c_0+b_1\zeta_n=0$, or equivalently a constant $\zeta_n$ and $\bar\zeta_n$ given as a linear function of $\zeta_{n-1}$. Since the support after replacing the monomials $\bar\zeta_n\zeta_{n-1}^a$ in $\bar f_{n-1}$ with $\zeta_{n-1}^a$ and $\zeta_{n-1}^{a+1}$ is also the same as the support of $\bar f_{n-1}$ projected onto the hyperplane of constant $\bar\zeta_n$, we actually have $V_{2n}(\dotsc,\Delta_n,I_n)=V_{2n}(\dotsc,\bar I_n,I_n)$ with the shear term present. Finally, with the symmetry argument $V_{2n}(\dotsc,\Delta_n,I_n)=V_{2n}(\dotsc,\bar\Delta_n,\bar I_n)$, we obtain
\begin{equation}\begin{split}
V_{2n}(\dotsc,F_n,\bar F_n)
&=(1+m_n^2+2m_n)V_{2n}(\dotsc,I_n,\bar I_n)
\\&=(1+m_n)^2V_{2n-2}(\dotsc,F_{n-1},\bar F_{n-1}),
\end{split}\end{equation}
so $V_{2n}(\dotsc,F_n,\bar F_n)=\prod_{j=1}^n(1+m_j)^2$ given $V_2(F_1,\bar F_1)=(1+m_1)^2$.

\end{appendix}

\label{finish}
\end{document}